\documentclass
[prl,amsfonts,superscriptaddress,twocolumn,showpacs,floatfix]{revtex4-1}%
\usepackage{amsmath}
\usepackage{amsfonts}
\usepackage{amssymb}
\usepackage{graphicx,graphics,color,epsfig}%

\begin{document}
\preprint{}
\title{Critical quasiparticle theory: Scaling, thermodynamic and transport properties}

\author{Elihu Abrahams}
\affiliation{Department of Physics and Astronomy, University of California Los Angeles, Los Angeles,
CA 90095}

\author{Peter W\"{o}lfle}
\affiliation{Institute for Theory of Condensed Matter and Center for Functional
Nanostructures, Karlsruhe Institute of Technology, 76131 Karlsruhe, Germany}\date{\today{}}

\begin{abstract}
We use the recently developed \cite{1} critical quasiparticle theory to derive the scaling behavior associated with a quantum critical point (QCP) in a correlated metal. This is applied to the magnetic-field induced QCP observed in YbRh$_2$Si$_2$ (YRS), for which we also derive the critical behavior of the  specific heat, resistivity, Gr{\" u}neisen coefficient, and the thermopower. The theory accounts very well for the experimental results.

\end{abstract}
\maketitle

\subsection{Introduction}
Recent advances in low-temperature experimental techniques have stimulated much interest in quantum critical phenomena, which comprise phase transitions at zero temperature (``quantum critical point") and associated effects due to quantum fluctuations at very low temperatures. Reference \cite{Piers} gives an introductory review  of the subject.

These developments have generated a variety of difficult theoretical questions; among them is the issue of how to treat the regime of strongly-interacting quantum fluctuations.
In a recent paper \cite{1}, we developed an extension ~of the quasiparticle concept of Fermi liquid theory to the non Fermi-liquid regime near a quantum critical point (QCP). In essence, the theory goes beyond the Gaussian regime of critical fluctuations by introducing interactions among the quantum fluctuations into the correlation function of the fluctuations. 
Central to the analysis is the concept of critical quasiparticles, which is based on the recognition that the single-particle spectral function can display a quasiparticle peak at non-zero excitation energy or temperature. This is expressed as a non-zero quasiparticle weight $Z(\omega)$ for $|\omega|$ not too small, although, as in a non Fermi liquid, at the Fermi surface $Z(\omega = 0)=0$.

We realized the critical quasiparticle theory for the case of an antiferromagnetic quantum critical point and applied it to several quantities, principally resistivity and specific heat, for successful comparison to experimental results on the heavy-fermion metal YbRh$_{2}$Si$_{2}$ (YRS), thereby showing that the theory, which describes a physically transparent scenario, is capable of accounting for experimental results on a quantum critical metal.

\begin{figure}[h]
\centering
\includegraphics[trim=8cm 0.7cm 2cm 3cm, clip=true, width=3.23in, angle=0]
{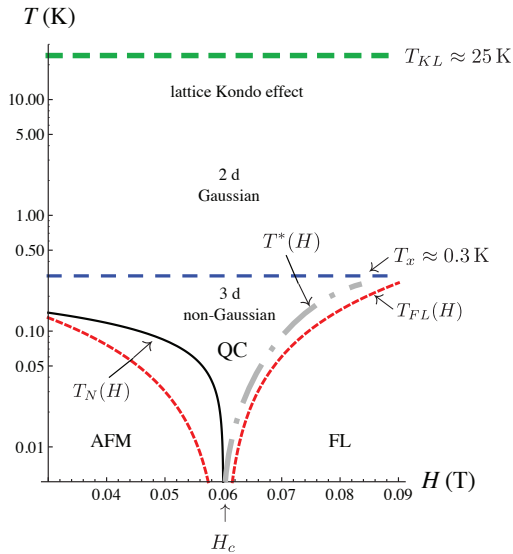}
%\vskip .8cm
\caption
{Phase diagram for YbRh2Si2 in the neighborhood of the critical magnetic field $H_c\simeq 0.06$ T: The  dashed lines represent crossovers. As the temperature is lowered below $T_{KL}$, the lattice Kondo effect and heavy quasiparticles develop and weakly-interacting (``Gaussian") 2d AFM fluctuations associated with the quantum critical point cause non-Fermi liquid behavior. The $T_x$ line represents the crossover to strongly-interacting 3d fluctuations, which are dominant within the cone of quantum criticality (QC), bounded on the right by the crossover at $T_{FL}(H)$ into the heavy Fermi-liquid state and on the left by the curved dashed line within the antiferromagnetic ordered state that sets in below $T_N(H)$. The $T^*(H)$ crossover (dot-dashed line) is discussed in the text, below Eq.\ (\ref{sol}).}
\label{phase}
\end{figure}

The implementation of the theory for an antiferromagnetic (AFM) quantum critical point for a heavy-fermion compound is based on the recognition that below a ``lattice Kondo temperature" $T_{KL}$, hybridization between conduction ($s,p,d$) electrons and local magnetic moments ($f$ orbitals) produces a heavy-electron liquid with an associated mass enhancement due to the originally localized character of the $f$ electrons. However, the $f$ electrons are also responsible for the antiferromagnetism in a region of the phase diagram of the material. Near the AFM critical point, critical spin fluctuations are enhanced and interact with the heavy quasiparticles. This produces further mass enhancement and occurs in two stages. Above a certain temperature $T_x$ but below $T_{KL}$, the critical fluctuations are Gaussian ({\it i.e} non-interacting); those of two-dimensional AFM character (or three dimensional ferromagnetic character) produce a logarithmic mass enhancement. A review of this physics can be found in reference \cite{LRVW}. Below $T_x$, however, the critical fluctuations begin to interact with each other and consequently the effects on the quasiparticles change, {\it e.g.} the effective mass, or $Z$-factor, acquires a singular power-law frequency dependence. This is the region described by the critical quasiparticle theory. Figure \ref{phase} illustrates these regions in the phase diagram for the heavy-fermion metal YRS.
In YRS, an AFM quantum critical point is accessed by tuning a magnetic field $H$, but the theory is appropriate whatever the nature of the ordered phase and whatever the tuning parameter.

Figure \ref{phase} shows the phase diagram for YbRh$_2$Si$_2$ near the quantum critical point. Five different regions may be seen: (1) The antiferromagnetically ordered region at temperatures below the N{\' e}el temperature $T_N(H)$; (2) The Landau Fermi-liquid regime at $T\lesssim T_{FL}(H)$; (3) A high temperature local-moment regime at $T\gg T_{KL}$, where $T_{KL}$ is the characteristic temperature of the lattice Kondo effect, below which coherent heavy quasiparticles form by hybridization of $f$-electrons and conduction electrons; (4) A regime of quantum critical quasi-two-dimensional antiferromagnetic Gaussian fluctuations  at $T_x\lesssim T\lesssim T_{KL}$, characterized by moderate non-Fermi liquid behavior; (5) The true critical regime, ``QC", for $T\lesssim T_x$ governed by three-dimensional antiferromagnetic fluctuations interacting strongly with the heavy quasiparticles.

The basic input for the theory is a phenomenological form for the critical antiferromagnetic (AFM) spin fluctuation correlator at low temperature below $T_x$, which reflects that it is generated from quasiparticles having a non-zero frequency-dependent $Z$-factor:
\begin{equation}
{\rm Im}\chi({\bf q},\omega)=\frac{(N_0/Z)(\omega/v_F^*Q)}{[r_0(H)+Z(\mathbf{q-Q})^{2}\xi_{0}^2]^2 +(\omega
/v_{F}^*Q)^2},
\label{chi}
\end{equation}
where $N_{0}$ is the bare density of states at the Fermi surface, $v_{F}%
^{\ast}= (m_b/m^*)v_F$ is the renormalized quasiparticle Fermi velocity, $\xi_{0}\simeq k_{F}^{-1}$ is
the microscopic AFM correlation length and $r_0(H)=1 + F(Q,H)\propto H/H_c - 1 $. Here $F(Q,H)$ is a dimensionless generalized Landau parameter,which $ \rightarrow -1$ at the critical point. For convenience, we denote the underlying tuning parameter by $H$, which for YRS is the magnetic field.
The interaction between quasiparticles and spin fluctuations $I$ is related to to $F(Q,H)$ by $I=Z F(Q,H)/N_0$. The ``bare" quantities $m_b,\, N_0, \, v_F$ here are to be understood as those quantities already renormalized at $T_{KL}$ by hybridization and the lattice Kondo effect, as discussed above.  The additional (frequency-dependent) mass enhancement caused by interaction with critical spin fluctuations is denoted as $m^*/m_b=1/Z$. We do not consider questions relevant to the crossover \cite{NPF} from the local moment phase at $T\gg T_{KL}$ to the heavy-fermion phase at $T\lesssim T_{KL}$. The experimentally determined values in YRS are $T_{KL}\approx 25$ K and $T_x \approx 0.3$ K.
%At non-zero temperature the control parameter acquires a shift which is expected to follow a power law: $r(H,T)=r(H,0)+r_T T^b$. The exponent $b$ may be determined from the magnetic field dependent N{\' e}el temperature $T_N=(H_c-H)^p$ (assuming symmetric behavior about $H_c$) as $b=1/p$. The published data on $T_N$ indicate a value of $p<1$.

The justification for the form of Eq.\ (\ref{chi}) is discussed in detail in reference \cite{1}. We neglect the momentum dependence of the $Z$-factor by assuming that the critical behavior at the hot spots is spread over the whole Fermi surface by impurity scattering or other interaction effects. The $Z$-factor is determined by the quasiparticle self energy, which in turn is determined by the interaction with the spin fluctuations.  This leads  \cite{1} to a self-consistency relation in the form of a differential equation for $Z(\omega)$:
\begin{equation}
Z^{-1}=1+\lambda \frac{d}{d\omega} (Z^{-3} \omega^{3/2}).
\label{self}
\end{equation}
This equation has two different physically meaningful solutions. The first is a weak coupling solution, valid provided the second term on the r.h.s. of Eq.\ (\ref{self}) is $\ll 1$ in the energy range considered. This corresponds to the conventional spin-density wave scenario as discussed in the works of Hertz \cite{weak1}, Millis \cite{weak2} and Moriya \cite{weak3}. There exists, however, a second solution in the strong-coupling domain; it is accessible provided the initial $Z^{-1}$ at the scale when the three-dimensional antiferromagnetic fluctuation regime is entered (in the case of YRS at $T\approx 0.3$K) is sufficiently large such that the second term on the r.h.s. of Eq.\ (\ref{self}) dominates. We conjecture that in YRS, two-dimensional antiferromagnetic (assisted by three-dimensional ferromagnetic) spin fluctuations above $T \approx 0.3$K can provide the necessary growth of the effective mass everywhere on the Fermi surface, even in the absence of impurity scattering. (In the case of quasi-2d antiferromagnetic fluctuations in a 3d metal this has been noticed first in reference \cite{rosch}). Then the strong-coupling solution is
\begin{equation}
Z^{-1}=3^{-1/2}(k_{F}\xi_{0})^{-3/2}(\omega/v_{F}Q)^{1/4}.
\label{sol}
\end{equation} 
At non-zero temperature, $Z$ is approximately given by $Z(\omega,T) \propto [\omega^{2}+(\pi T)^{2}]^{1/8}$.   

In the following, we describe the phase diagram obtained within our theory as consisting of the ordered phase bounded by the N{\' e}el temperature $T_N(H)$, the critical regime (``critical cone") at $T_x>T>T_{FL}(H)$ and the Landau Fermi liquid regime at $T<T_{FL}(H)$ (see figure \ref{phase}). In addition to these phase boundaries and crossover lines, 
Hall effect and other measurements in YRS \cite{hall1,hall2,science} show anomalies at a temperature $T^*(H)$ in the critical region. The origin of the $T^*(H)$ line is not clear at present. Recent experiments on YRS under pressure \cite{tokiwa1} or YRS doped with Co or Ir \cite{fried} have shown that while the critical field for the magnetic transition is found to shift, the $T^{*}$ line stays unchanged. This suggests that the $T^{*}$ feature is only weakly or not at all tied to the AFM transition. The present theory ascribes the critical behavior to interaction of renormalized quasiparticles with renormalized antiferromagnetic spin fluctuations, rather than with fluctuations associated with $T^*(H)$, as has been suggested elsewhere \cite{science,si1,si2}.

In this paper, we elaborate the critical quasiparticle theory further by deriving the scaling behavior of quantities near the QCP and we add the critical behavior of the Gr{\" u}neisen coefficient, thermopower, thermal expansion, magnetization, and susceptibility to the experimental consequences of the theory. Again, we compare these results of the theory to experimental information on YbRh$_{2}$Si$_{2}$, which has a magnetically tuned QCP.

The deduction of the critical behavior of several thermodynamic quantities from the scaling form of the free energy has been discussed by Zhu, {\it et al} \cite{Zhu}. The corresponding exponents for these quantities are determined by the correlation length exponent $\nu$, the dynamical exponent $z$ and the dimensionality $d$.

\subsection{Scaling} As discussed in reference \cite{1}, the critical regime is characterized by fluctuations of dimensionality $d=3$. We can determine $\nu$ and $z$ from the critical quasiparticle theory as follows: It can be shown that when the frequency dependence of $Z$ is a power less than $1/2$, the frequency-dependent $Z$-factors appearing in Eq.\ (\ref{chi}) have the argument $\omega$. Then from Eq.\ (\ref{chi}) and the result $Z(\omega)\propto \omega^{\alpha}$,  where we found $\alpha=1/4$, we can determine the dynamical exponent $z = 4$ since at the critical point
\begin{equation} 
{\rm Im}\chi({\bf q},\omega)^{-1} \sim Z^2(\omega)q^4 + \omega^2/Z^2(\omega).
\label{chiqp}
\end{equation}
Here and in the following, ${\bf q}$ denotes the deviation from the ordering wave vector ${\bf Q}$.
The scaling exponent of the spatial correlation length $\xi$ is found as $\nu = 1/3$ from 
$
r_0(H) \sim Zq^2 \sim (q^z)^{1/4}q^2 \sim q^3
$
so that $\xi \sim r_0^{-1/3}$. At $T=0$ we thus find $\xi \propto (H-H_c)^{-1/3}$.% whereas at $H=H_c$, we have $\xi \propto T^{-b/3} \sim (H-H_c)^{-1/3}$.
 The characteristic correlation time is given by $\tau_c \sim \xi^z \sim (H-H_c)^{-4/3}$.
  
When $H=H_c$ ({\it i.e.} $r_0=0$), the approach to the $T=0$ critical point is described by a temperature dependence of the control parameter, which can be written $r(H_{c},T) \sim  T^{1/z\nu}$. The condition $r(H_{c},T) \approx r(H,T=0)$  or $T=T_{FL}(H) \propto |H-H_c|^{z\nu}$ defines the boundary of  the ``critical cone" of  quantum criticality.
Then, within the quantum critical regime, the temperature dependence of the correlation length is found as $\xi \propto T^{-1/z}$. It may be shown that the above critical temperature dependent correction to $r$ is caused by the interaction of spin fluctuations treated at the Hartree level \cite{PW}, giving rise to $r(H_{c},T) \sim  T/Z(T)$.

From these results ($\nu = 1/3,\, z=4$), we may write ${\rm Im}\chi({\bf q},\omega)$, or equivalently the structure factor $S(q,\omega)$, in the scaling form
\begin{equation}
S(q,\omega) = \xi^4\Phi(\omega\xi^4,q\xi), \
\Phi(x,y) \propto \frac{x^{1/2}}{(1+x^{1/4}y^2)^2 + x^{3/2}}.
\end{equation}
Note that at the critical wave vector (at $q=0$), $\omega/T$ scaling holds.

The  scaling form for the free energy density may be deduced by dimensional analysis \cite{LRVW}:
\begin{subequations}
\begin{align}
f(H,T) &= \xi^{-(d+z)}\Phi_f(r_0\xi^{1/\nu}, T\xi^z)\nonumber \\
&\rightarrow T^{\frac{d+z}{z}} \Psi\left(\frac{r_0}{T^\frac{1}{z\nu}}\right) \label{scalea}\\
&{\rm or} \;\; r_0^{\nu(d+z)}\widetilde\Psi\left(\frac{T}{r_0^{z\nu}}\right)\label{scaleb}.
\end{align}
\end{subequations}
In our case, $d=3,\;\;\nu=1/3,\;\;z=4$.

We may generalize these results to arbitrary dimension $d$. First, we note (from the derivation of the quasiparticle weight factor $Z$ presented in reference 1) that $d=4$ is the upper critical dimension, so that the spin fluctuations do not destroy the Fermi liquid state for $d>4$. At $d<2$, we find that the critical quasiparticles are no longer well defined. We therefore consider here only dimensions within the interval $2\leq d\leq 4$.
Using the exponent $\alpha$ to describe the frequency power law of the quasiparticle weight $Z$, we find, from Eq.\ (\ref{chiqp}), that the dynamical critical exponent is given by $z=2/(1-2 \alpha)$ and the correlation length exponent is $\nu=1/(2+z \alpha)$
%We find that the quasiparticle weight varies as $\omega^{(d/2-1)/(d-1)}$. Consequently, both exponents $z$ and $\nu$ are independent of $d$.

\subsection{Specific heat} The critical part of the specific heat at the critical tuning parameter $r=0$ is found from Eq.\ (\ref{scalea}):
$C_c\propto T^{d/z} = T^{3/4}$ as we found before \cite{1}. We can obtain the behavior of the specific heat near $T=0$ as a function of $r$ from Eq.\ (\ref{scaleb}) as follows \cite{Zhu}: Near $x=0$, $\widetilde\Psi(x)$ behaves as $\widetilde\Psi(x\rightarrow 0) = \widetilde\Psi(0) + bx^\mu$. Therefore \begin{align}
C(r, T) & \propto r^{(d+z)\nu -\mu z\nu}\mu(\mu -1) T^{\mu -1}\nonumber \\
& = \mu(\mu -1) T^{\mu -1} r^{-1/3(4\mu - 7)}.
\label{C}
\end{align}
This behavior is valid in the region of the $T,H$ phase diagram defined by $T/r^{z\nu} < 1$, which is $T<T_{FL}(H) \propto |H-H_c|^{z\nu}$.

As we showed in reference \cite{1}, the theoretical behavior of $\gamma = C_c/T$ at criticality found above agrees very well with the data. For the low $T$-dependence of $\gamma$ on $r= |H-H_c|$, we note that in YRS on either side of the QCP, Fermi liquid behavior obtains \cite{custers} so that $\mu =2$. Then, from Eq.\ \ref{C}, $\gamma = C/T$ varies as $|H-H_c|^{-1/3}$. This is precisely the behavior observed in experiment on YRS \cite{custers,hartmann}.

\subsection{Resistivity} The scaling analysis may be applied to determine the temperature and magnetic field power-law dependences of the resistivity, which (assuming impurity and umklapp scattering) is given by:
\begin{equation*}
\rho(T,H)-\rho(0) \propto \Gamma(T,H)/Z(T,H),
\end{equation*}
where $\Gamma$ is the quasiparticle relaxation rate, which we found \cite{1} at $d=3$ and $H=H_c$, to be proportional to $\omega^{3/2}/Z^2(\omega)$. Since $\Gamma$ has dimensions of energy, its scaling form is
$\Gamma(T,H) = \xi^{-z}\Phi_{\Gamma}(T\xi^z)$ as in Eq.\ (\ref{scaleb}).
With $Z(\omega)\propto\omega^{\alpha}\propto \xi^{-z\alpha}$, we have
\begin{equation*}
\rho(T,H)-\rho(0) \propto \xi^{(\alpha -1)z}\Phi_{\Gamma}(T\xi^z)
\end{equation*}
To insure that $\rho$ is finite at $H=H_c$ and $T > 0$, we require $\Phi_{\Gamma}(x) \propto x^{1-\alpha}$ as $x\rightarrow \infty$, so that
\begin{equation}
\rho(T,H_c)-\rho(0) \propto T^{1-\alpha}.
\end{equation}

As we discussed above for the specific heat, away from criticality, at low temperature outside either side of the critical cone defined by $T= T_{FL}$, Fermi liquid behavior, $\rho(T)-\rho(0) \propto T^2$ obtains, so that $\Phi_{\Gamma}(x) \propto x^2$. Then
\begin{equation}
\rho(T,H)-\rho(0) \propto \xi^{(\alpha+1)z}T^2.
\end{equation}
Using the values $\alpha=1/4, z=4$, which we found in the self-consistent critical quasiparticle theory \cite{1}, we find that Eqs.\ (6,7) become
\begin{subequations}
\begin{align}
\rho(T,H_c)-\rho(0) & \propto T^{3/4} \label{rhoa}\\ 
\rho(T,H)-\rho(0) & \propto \xi^5 T^2 \propto |H-H_c|^{-5/3}T^2.\label{rhob}
\end{align}
\end{subequations}
The available data agree very well with Eq.\ (\ref{rhoa}) as found in reference \cite{1} by direct calculation. At the time of this writing, for the field dependence of the coefficient of $T^2$ in  Eq.\ (\ref{rhob}), there is insufficient data close enough to $H_c$ to allow a quantitative comparison.

\subsection{Thermopower}
At very low temperatures, when impurity scattering
dominates, the thermopower $S$ may be expressed in terms of the electrical
conductivity $\sigma _{imp}(\mu )$, depending on the chemical potential $\mu $, by the Mott formula  
\begin{equation}
S=\frac{\pi ^{2}}{3}\frac{T}{e}\frac{\partial \ln \sigma _{imp}}{\partial
\mu }.
\end{equation}%

%%%%%Fig Thermo

\begin{figure}[h]
\centering
\includegraphics[trim=0cm 2cm 0 1cm, clip=true, width=2.9in, angle=-90]
{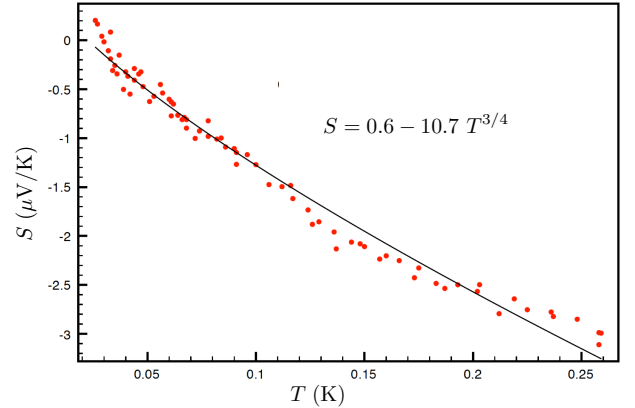}
\vskip -.8cm
\caption
{Thermopower: Comparison of theory and data of reference \cite{hartmann} near the critical magnetic field and below the critical temperature for quantum critical scaling.}
\label{thermo}
\end{figure}

Since the mean free path $\ell$ is not renormalized, $\sigma _{imp}(\mu)\propto k_F^2\ell$, where $k_F$ is the Fermi wave vector, is independent of the effective mass. The derivative with respect to $\mu$ may be
expressed by the one with respect to the Fermi wave vector as $\partial
/\partial \mu =(1/v_{F}^*)(\partial /\partial k_{F})$ and $
\partial \sigma _{imp}/\partial k_{F}$ is independent of $m^{\ast }$. Therefore, we find 
$S\propto Tm^{\ast }(T)$. In the Gaussian fluctuation regime, we get $%
S\propto T\ln (T)$, which has been observed \cite{hartmann}. In the critical
regime proper, at the critical field $H=0.064$ T we predict $S\propto T^{3/4}$. The comparison of our result to the experiment reported in reference \cite
{hartmann} is shown in figure \ref{thermo}. Unfortunately, the data was taken at $H=0.06$ T, below the critical field of 0.064 T. At the measurement field,  the AFM N{\' e}el temperature $T_N$ is about 20 mK \cite{hartmann}, so we do not expect good agreement at the very lowest temperatures, as seen in the figure.

\subsection{Magnetization and susceptibility} 
\begin{figure}[h]
\centering
\includegraphics[trim=1cm 2.5cm .2cm 2.5cm, clip=true, width=2.8in, angle=-90]{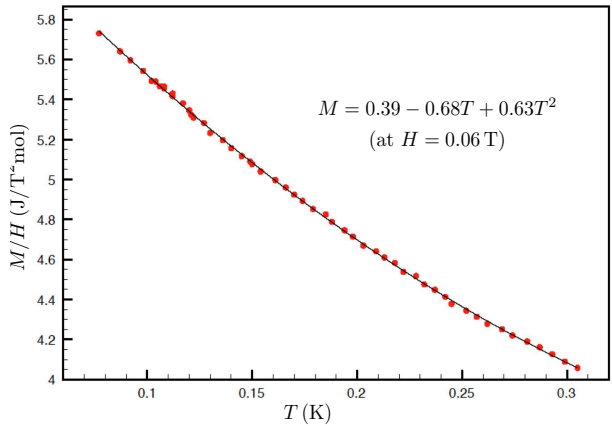}
\vskip -.8cm
\caption
{Magnetization: Comparison of theory and data of reference \cite
{Tokiwa} near the critical magnetic field. A $T^2$ Fermi-liquid contribution has been added to the critical behavior.}
\label{mag}
\end{figure}

The temperature dependence of the magnetization may be found from the derivative of the free energy in Eq.\ (\ref{scalea}) with respect to field:
\begin{align}
M_{{\rm crit}} &= -\left(\frac{\partial f}{\partial H}\right)_T = -\frac{1}{H_c}\left(\frac{\partial f}{\partial r_0}\right)_T \nonumber\\
&=-\frac{1}{H_c}T^{\frac{d+z}{z}}\Psi '\left(\frac{r_0}{T^{\frac{1}{z\nu}}}\right)T^{-\frac{1}{z\nu}}.
\label{mac}
\end{align}
Assuming $M$ is regular at $r_0=0$ so that $\Psi'(0) = {\rm const}$, we get
\begin{equation}
M(T) - M(0) \propto -T^{\frac{d+z-1/\nu}{z}} = -T.
\label{M(T)}
\end{equation}
This result agrees fairly well with the experimental data at low $T$ (below 400 mK) as reported in reference \cite%\onlinecite
{Tokiwa}. A very good fit is obtained if the critical $T$-dependence of Eq.\ (\ref{M(T)}) is augmented by the sub-leading $T^2$ contribution due to Fermi-liquid effects. This is shown on figure \ref{mag}, where the theoretical behavior at $H=0.06$\,T is given by
\begin{equation}
M(T) - M(0) = -0.68 T + 0.63T^2.
\label{MFL}
\end{equation}

%%%%%%Fig Mag

For the field dependence at small $T$, we use Eq.\ (\ref{scaleb}):
\[
M_{{\rm crit}} =-\left(\frac{\partial f}{\partial H}\right)_{T=0}= -\frac{1}{H_c}\nu(d+z)r_0^{\nu(d+z)-1}{\widetilde\Psi}(0),
\]
which for $d=3,\, z=4, \, \nu =1/3, \, r_0 = H/H_c - 1$ is
\begin{equation}
M(H) - M(H_c) = -\frac{7}{3}\frac{1}{H_c}\left(\frac{H}{H_c} - 1\right)^{4/3}\widetilde{\Psi}(0).
\end{equation} 
Therefore, the critical field dependence of the susceptibility is
\begin{equation}
\chi_{{\rm crit}}= \frac{\partial M}{\partial H} \propto -(H-H_c)^{1/3}.
\end{equation}
The available data \cite{Tokiwa2} indicate a cusp-like behavior of $\chi(H)$ when approaching $H_c$, but do not extend close enough to allow for a quantitative comparison.

\begin{figure}[h]
\centering
\includegraphics[trim=2.3cm 2.7cm 0cm -3cm, clip=true, width=2.8in, angle=-90]{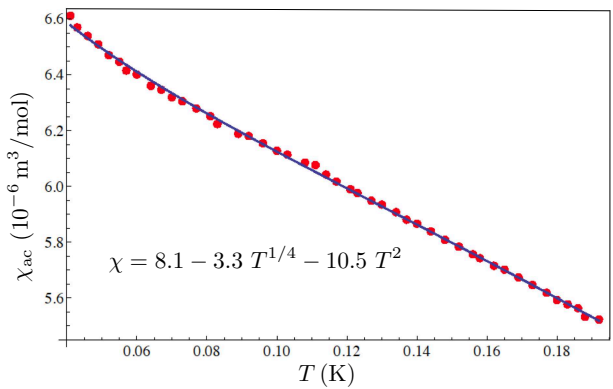}
\vskip -1.4cm
\caption
{Magnetic susceptibility: Comparison of theory, Eq.\ (\ref{chiT}) and data of reference \cite
{gegen} near the critical magnetic field and below the critical temperature for quantum critical scaling. Similarly to Eq.\ (\ref{MFL}) and figure \ref{mag}, we have added a $T^2$ Fermi-liquid contribution.}
\label{gegen}
\end{figure}

The critical $T$-dependence of the susceptibility is found from Eq.\ (\ref{mac}) as
\begin{equation}
\chi = \frac{\partial M}{\partial H}= \frac{1}{H_c}\frac{\partial M}{\partial r_0} \propto T^{\frac{d+z-2/\nu}{z}} \propto T^{1/4},
\label{chiT}
\end{equation}
where we assumed $\Psi''(0) = {\rm const}$. We can compare this result to the experiment reported by Gegenwart {\it et al} \cite{gegen}, where $\chi_{\rm ac}$ was measured below 0.2 K at a field of 0.05 T. Since $T_N$ is about 30 mK at this field, we do not include data below 40 mK in the comparison shown on figure \ref{gegen}.

%%%Fig gegen

\subsection{Gr{\" u}neisen ratio} In the case of a magnetic field tuned QCP, as in YRS, the magnetic Gr{\" u}neisen ratio \cite{Zhu} is studied:
\begin{equation}
\Gamma_M = - \frac{(\partial M/\partial T)_H}{C_H} = -\frac {(\partial S/\partial H)_T}{C_H}
\label{gru}
\end{equation}
The critical part of the Gr{\" u}neisen ratio has been discussed in reference \cite{Zhu} and measured in YRS by
Tokiwa, {\it et al} \cite{Tokiwa}. As in reference \cite{Zhu}, the behavior of $\Gamma_M$ may be determined from the scaling analysis. We find, for $z=4$, $\nu = 1/3$ and $\mu =2$, that $[\partial S/\partial H]_{r_0=0}\propto T^0$. Then
\begin{subequations}
\begin{align}
\Gamma_M(r=0,T) & \propto T^{-3/4} \label{GT}\\
\Gamma_M(r, T\rightarrow 0) & = \frac{-1/3}{H-H_c}\label{Gr}.
\end{align}
\end{subequations}
The factor $-1/3 =\nu(d-z)$ in Eq.\ (\ref{Gr}) is universal as the scaling functions in the ratio defining $\Gamma_M$ cancel.

These results as well as those for the thermodynamic Gr{\" u}neisen parameter can also be obtained \cite{PW} by direct calculation from derivatives of the self-energy expressions given in reference \cite{1}.

\begin{figure}[h]
\centering
\includegraphics[trim=0cm 2cm 0 1cm, clip=true, width=2.9in, angle=-90]{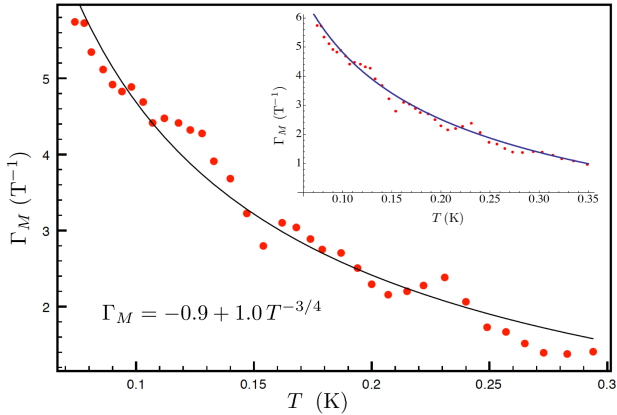}
\vskip -.8cm
\caption
{Magnetic Gr{\" u}neisen ratio: Comparison of theory, Eq.\ (\ref{GT}) and data of reference \cite
{Tokiwa} at the critical magnetic field and below the critical temperature for quantum critical scaling. The inset shows the comparison of the data and Eq.\ (\ref{gru}) when the previously fit $M(T)$ and $C(T)$ are used.}
\label{grun}
\end{figure}

Just as for the specific heat and the resistivity \cite{1}, there is excellent agreement between the critical quasiparticle theory and the experiment \cite{Tokiwa} for the magnetic Gr{\" u}neisen ratio. The factor $\nu(d-z)=-1/3$ in Eq.\ (\ref{Gr}) is measured as $-0.30\pm 0.01$ and the comparison of our result of Eq.\ (\ref{GT}) of the $T^{-3/4}$ dependence at $H=H_c$ to the experiment is shown on figure \ref{grun}. 

We may evaluate the consistency of our results by using the theory fits for $M(T)$, Eq.\ (\ref{MFL}) and for the specific heat shown on figure  1 of  reference \cite%\onlinecite
{1} ($C= 0.034 + 0.454 T^{3/4}$) in Eq.\ (\ref{gru}). There are no further adjustable parameters and the agreement is shown in the inset to figure \ref{grun}.

The volume thermal expansion coefficient $\beta(T)$ enters the conventional thermodynamic Gr{\" u}neisen ratio $\Gamma_T = \beta(r_0, T)/C(r_0,T)$. The critical behavior here is tuned by pressure $P$, so that $r_0 \propto (P_c - P)/P_c$. The critical part of $\beta$ is determined by derivatives of the free energy as
\[
\beta_c(r_0, T) = \frac{1}{Pv}\frac{\partial^2 f}{\partial T \partial r_0} \propto T^0,
\]
where $v$ is the specific volume and we have used Eq.\ (\ref{scalea}). To this we may add the conventional Fermi-liquid contribution $\propto T$ so that $\beta = a + bT$ in the critical region.

\subsection{Conclusion}~  ~The critical quasiparticle theory describes a renormalized Gaussian picture of critical fluctuations, where the fluctuations are assumed to be non-interacting beyond effects that renormalize the quantities entering the fluctuation spectrum. The renormalizations of these parameters, the staggered  static susceptibility and the Landau damping term, are, however, of a strong-coupling nature. It remains to be shown that fluctuation interactions are irrelevant within a more complete theory. We have used a scaling analysis of the  theory to derive the critical behavior of a number of experimentally observed quantities in the heavy fermion compound YbRh$_2$Si$_2$. The good agreement with all these observed quantities indicates that the critical quasiparticle theory captures essential features of the critical behavior in this compound. We note that YbRh$_2$Si$_2$ is especially suitable for application of the theory since it has the wide region of two-dimensional (2d) critical behavior that is necessary to ultimately access the strong coupling regime of three-dimensional fluctuations at lower temperatures, as discussed in reference \cite{1}, where  the critical quasiparticle theory was introduced.

Besides YRS, other candidate systems have been studied with different degrees of detail. For CeCu$_{6-x}$Au$_x$ it has been found (for a review see reference \cite{LRVW}) that some of the low-temperature properties like the specific heat and the resistivity indicate quasi-2d antiferromagnetic fluctuations in the Gaussian regime. Neutron scattering studies at somewhat higher energies have been interpreted as showing critical scaling behavior incompatible with the Gaussian theory. It is conceivable that the behavior is different depending on the energy scale. Further candidate systems like CeCoIn$_5$ also appear to show quasi-2d antiferromagnetic Gaussian fluctuations (for an overview, see reference \cite{LRVW}). In all of these systems we expect a crossover to 3d fluctuations at lower temperatures, in which case our theory should apply. 

\begin{acknowledgments}

We thank J{\" o}rg Schmalian for valuable suggestions and Alexander Balatsky, Andrey Chubukov, Max Metlitski, Subir Sachdev, and Qimiao Si  for useful discussions. This work was supported in part by the
DFG research unit 960 "Quantum phase transitions". Part
of this work was carried out at the Aspen Center for Physics (NSF grant 1066293),  Los Alamos National Laboratory and as a Carl Schurz Memorial Professor at the University of Wisconsin, Madison (PW). We thank these institutions for their hospitality and support.
\end{acknowledgments}

\end{document}